\documentclass[11pt]{article}
\usepackage{moriond,epsfig,xspace}

\def\Journal#1#2#3#4{{#1} {\bf #2}, #3 (#4)}

\def\PRL{\em Phys. Rev. Lett.}
\def\PRD{{\em Phys. Rev.} D}

\def\etal {{\em{at al.}}\xspace}

\def\be{\begin{equation}}
\def\ee{\end{equation}}
\def\bea{\begin{eqnarray}}
\def\eea{\end{eqnarray}}

\def\iab{\ensuremath{\mathrm{ab}^{-1}}\xspace}
\def\ifb{\ensuremath{\mathrm{fb}^{-1}}\xspace}

\def\squark{\ensuremath{\tilde{q}}\xspace}
\def\stop{\ensuremath{\tilde{t}}\xspace}

\def\CP{{\rm CP}\xspace}
\def\nubar{\ensuremath{\overline{\nu}}\xspace}

\usepackage{relsize}
\def\babar{\mbox{\slshape B\kern-0.1em{\smaller A}\kern-0.1em
    B\kern-0.1em{\smaller A\kern-0.2em R}}\xspace}

\begin{document}
\vspace*{4cm}
\title{SuperB}

\author{ A.~J.~BEVAN }

\address{Department of Physics, Queen Mary University of London, Mile End Road, London E1 4NS, England}

\maketitle\abstracts{ An overview of the SuperB project and its physics programme is presented.  There are many new
physics sensitive observables that can be measured at a high luminosity $e^+e^-$ collider operating near a centre of
mass energy of $\sim 10$ GeV, and many new physics scenarios to test in the literature.  Together these form a {\em
golden matrix} of observables versus scenarios.  Each scenario has its it's own golden channel(s) and the pattern of
deviations from Standard Model behaviour that will be measured by SuperB can be used to home in on the new physics
scenario describing nature at high energies. }

\section{Introduction}

Two related problems facing modern physics are the universal asymmetry problem and the flavour problem. The universal
asymmetry problem arises from the fact that the known level of CP violation as described by the
Cabibbo-Kobayashi-Maskawa (CKM) quark mixing matrix \cite{bevan:cabibbo,bevan:km} is a billion times too small to
account for our matter dominated universe.  The flavour problem is the tension between our expectations that new
physics (NP) scale $\Lambda_{\mathrm{NP}}$ should occur at a TeV energy scale and our understanding that flavour
changing neutral currents place strong constraints on NP scenarios and impose much higher scales for
$\Lambda_{\mathrm{NP}}$.  Precision electroweak constraints imply $\Lambda_{\mathrm{NP}} \sim 1$ TeV, yet at the same
time flavour changing neutral current measurements require $\Lambda_{\mathrm{NP}} \geq 10-100$ TeV.

\section{SuperB in a nutshell}

SuperB is a next generation high luminosity asymmetric energy $e^+e^-$ collider proposed to be built on the campus of
Tor Vergata University of Rome.  The purpose of this experiment will be to elucidate the nature of NP in a way that
compliments the existing direct search programme at CERN. The accelerator will have 7 GeV electrons colliding with 4
GeV positrons, with a 15mrad crossing angle.  The Lorentz boost of the resulting center of mass system is approximately
half of that at the SLAC B-factory. The finite beam crossing angle results in a lower background that the current
B-factories. To compensate for the geometrical effect of the crossing, sextupole magnets will be used before and after
the interaction point (IP) in the so-called {\em crabbed waist} scheme in order to maintain maximal overlap of
colliding bunches~\cite{bevan:crabbedwaist}. The design luminosity for the accelerator is $10^{36}\, \mathrm{cm}^{-2}
\mathrm{s}^{-1}$, with a total of 75\iab of data being delivered at the $\Upsilon(4S)$ resonance in the first five
years of operation. To achieve this luminosity the vertical beam-size will be of the order of 20nm, so the collider
will operate with a small vertical emittance.  In addition to operating at the $\Upsilon(4S)$ centre of mass (CM)
energy, the accelerator will be able to operate at other energies including near charm threshold at the $\psi(3770)$
and at the other $\Upsilon$ resonances.

It is anticipated that SuperB will reuse a number of components of the SLAC B-factory, including parts of the PEP-II
accelerator complex, the super-conducting solenoid magnet, barrel of the electromagnetic calorimeter, and the quartz
bars of the detector of internally reflected cerenkov radiation (DIRC).  There are a number of proposed improvements to
the detector technology compared to the \babar detector, including faster readout, a smaller stand-off-box for the
SuperB DIRC, a layer 0 of Monolithic Active Pixel Sensors, and a muon system similar to the MINOS scintillating fibre
detector. The SuperB project and its physics goals are also described at length in a Conceptual Design Report
\cite{bevan:cdr} and the proceedings of a Valencia Physics Workshop \cite{bevan:valencia}. There is a great deal of
international interest in SuperB and a Technical Design Report (TDR) describing the details of the SuperB detector,
accelerator and physics goals is in preparation.  This is expected to be completed within the next two years.  The
remainder of these proceedings concentrate on the physics potential of SuperB.

\section{Physics potential of SuperB}

\subsection{Measurements sensitive to new physics}

\subsubsection{$\tau$ decays}

The search for Lepton Flavour Violation (LFV) in $\tau$ decay is complementary with the existing neutrino experiments
aimed at measuring $\theta_{13}$ and the MEG experiment at PSI that is dedicated to the search for $\mu\to e \gamma$.
Table~\ref{tbl:taulfv} summarizes anticipated sensitivities for several $\tau$ LFV searches possible at SuperB.  SuperB
has a 85\% polarization for the electron beam which is instrumental in reducing backgrounds in searches and results in
a better than $\sqrt{N}$ scaling from the current B-factories.  The sensitivities achievable are between 10 and 50
times more stringent than possible enhancements from new physics.  SuperB has the potential to severely constrain or
discover new physics in these scenarios. Recent sensitivity studies from the LHC experiments indicate that the LHC will
not be competitive with SuperB for any of these channels~\cite{bevan:lhclfv}. SuperB has a better search capability
than SuperKEKB project being pursued in Japan as SuperB uses polarized beams (leading to reduced backgrounds) and has a
larger target luminosity than SuperKEKB.

\begin{table}[!ht]
\caption{The experimental sensitivities (in units of $10^{-10}$) expected for LFV searches in $\tau$
decay.}\label{tbl:taulfv}
\begin{center}
\begin{tabular}{l|ccc}\hline
 Final State     & Sensitivity $/ 10^{-10}$ \\ \hline \hline
 $\mu\gamma$     & 20 \\
 $e\gamma$       & 20 \\
 $3\mu$          & 2  \\
 $3e  $          & 2  \\
 $\mu\eta$       & 4  \\
 $e\eta$         & 6  \\
 $\ell K^0_S$    & 2  \\
\end{tabular}
\end{center}
\end{table}

\subsubsection{Rare decays}

SuperB will be able to measure many rare decay processes that are sensitive to different scenarios of new physics. Each
scenario has its own golden channels, together they form a {\em golden matrix} of observables versus models.  The
golden channels are clear signals for new physics in one or more scenarios and by measuring the set of observables for
all of these golden channels it is possible to distinguish between many different types of new physics. This pattern
matching process using rare decays (and the processes described in other sections of these proceedings) is required as
it will not be possible for the LHC experiments to determine the precise nature of new physics through its extensive
programme of measurements.  Table~\ref{tbl:goldenmatrix} summarises the golden matrix for rare $B$ decays. In most
cases SuperB will be able to measure the observables listed in this matrix to the few percent level.  In particular
with 75\iab of data SuperB's measurement of ${\cal B}(B\to \tau \nu)$ will have a new physics search potential above
1TeV, and will be able to observe the decay $B\to K\nu\overline{\nu}$ occurring at the expected SM rate.  The expected
experimental sensitivities of these observables is listed in Table~\ref{tbl:goldenmatrixmeas}.

\begin{table}[!ht]
\caption{The golden matrix of observables versus new physics scenarios. L denotes a large effect, M denotes a
measurable effect, and CKM denotes a measurement that also requires precision determination of the CKM
matrix.}\label{tbl:goldenmatrix}
\begin{center}
\small
\begin{tabular}{l|ccccc}\hline\hline
                                  & $H^+$              & MFV  & Non-MFV & NP         & Right-handed \\
                                  & high $\tan \beta$  &      &         & Z-penguins & currents     \\ \hline
${\cal B}(B\to X_s\gamma)$        &                    &  L   &    M    &            &      M       \\
${\cal A}_{CP}(B\to X_s\gamma)$   &                    &      &    L    &            &      M       \\
${\cal B}(B\to \tau \nu)$         & L-CKM              &      &         &            &              \\
${\cal B}(B\to X_s\ell\ell)$      &                    &      &    M    &     M      &      M       \\
${\cal B}(B\to K\nu\overline{\nu})$ &                  &      &    M    &     L      &              \\
$S_{K_S\pi^0\gamma}$               &                   &      &         &            &      L       \\
The angle $\beta$ ($\Delta S$)    &                    &      &   L-CKM &            &      L       \\ \hline
\end{tabular}
\end{center}
\end{table}

\begin{table}[!ht]
\caption{Experimental sensitivities of the observables in the golden matrix shown in Table~\ref{tbl:goldenmatrix}. Here
X denotes that this measurement is not possible with a given data sample.}\label{tbl:goldenmatrixmeas}
\begin{center}
\small
\begin{tabular}{l|ccccc}\hline\hline
                                    & Current      & 10\iab  & 75\iab \\ \hline
${\cal B}(B\to X_s\gamma)$          &   7\%        &  5\%    &  3\% \\
${\cal A}_{CP}(B\to X_s\gamma)$     &   0.037      &  0.01   &  $0.004-0.005$    \\
${\cal B}(B\to \tau \nu)$           &   30\%       &  10\%   &  $3-4\%$    \\
${\cal B}(B\to \mu \nu)$            &   X          &  20\%   &  $5-6\%$    \\
${\cal B}(B\to X_s\ell\ell)$        &   23\%       &  15\%   &  $4-6\%$    \\
${\cal B}(B\to K\nu\overline{\nu})$ &   X          &  X      &  $16-20\%$    \\
$S_{K_{S}\pi^{0}\gamma}$            &   0.24       &  0.08   &  $0.02-0.03$    \\
The angle $\beta$ ($\Delta S$)      &   0.07       &  0.02   &  0.01    \\ \hline
\end{tabular}
\end{center}
\end{table}

\subsubsection{SUSY CKM}

As an illustration of the power of rare $B$ decays in constraining new physics we consider constraints on the
parameters of the SUSY CKM sector.  In MSSM with right handed neutrinos one has to measure 160 parameters in order to
fully constrain the theory.  Of these 160 parameters, 110 are flavour changing parameters.  So over two thirds of the
information required in order to fully constrain MSSM is the domain of flavour physics measurements.  We know that
quarks and neutrinos mix in a non trivial way, the latter introducing LFV to the SM. It follows that any natural
extension of the SM introducing super-parters of the quarks and leptons will have a non-trivial set of flavour
couplings that need to be determined.  As new CP violating phases could be introduced by new physics it is necessary to
measure the real and imaginary parts of the flavour couplings which can be denoted $(\delta_{ij}^q)_{AB}$, where $q$ is
a quark, $i$ and $j$ are squark indices and $AB$ are different combinations of left and right handed
currents~\cite{bevan:susyflavour}. The magnitude of these couplings is inversely proportional to the square of the
squark mass $m^2_{\squark}$ in the effective Lagrangian describing new physics. The corollary of this is that a
measurement of either the magnitude or the real and imaginary parts of $(\delta^q_{ij})_{AB}$ will constrain
$m_{\squark}$.

For example, with a data sample of 75\iab and using the branching fractions of inclusive $b$ to $s\ell\ell$ and $s
\gamma$ processes and the direct \CP asymmetry measurement of $b\to s\gamma$, SuperB will be able to make a percent
level measurement of the real and imaginary parts of $(\delta^d_{23})_{LR}$ or rule out a squark signal at the TeV
scale. The contribution to these proceedings by G.~Hiller discusses the possibility of using an extremely long lived
\stop to measure some of the off-diagonal SUSY CKM parameters at energy frontier machines~\cite{bevan:hiller}. However
if the \stop is short lived then these machines will only be able to measure the diagonal terms in this matrix. In
either case, SuperB is able to constrain many of the off-diagonal terms irrespective of the \stop lifetime thus
complementing the physics programme at the LHC. Examples of the SuperB capability to constrain the SUSY-CKM sector can
be found in~\cite{bevan:cdr,bevan:valencia}.

\subsubsection{Light Higgs Searches}

The decays $\Upsilon(NS) \to \ell\ell$ where $N=1,2,3$, and $\ell$ is a charged lepton, can be used to test lepton
universality (LU)~\cite{bevan:sanchis}.  In the SM the coupling constants of the different generations of leptons are
the same, so ratios of branching fractions of $\Upsilon$ decays to $e^+e^-$, $\mu^+\mu^-$, and $\tau^+\tau^-$ can be
compared to one another in order to search for violation of LU.  One possible mechanism that could violate LU is a
light neutral Higgs particle $A^0$ with a mass $\sim 10$ GeV. Many popular scenarios of NP include such a particle, for
example 2HDM and NMSSM~\cite{bevan:variouslumodels,bevan:mcelrath}. So if in addition to the SM amplitude, there is a
contribution to the $\Upsilon(NS) \to \ell\ell$ decay from a state with an intermediate $A^0$ this might be detectable
through the observation of LU violation at SuperB.

\subsubsection{Dark Matter Searches}

There are many possible dark matter scenarios, however to date there is no confirmed evidence for a dark matter
candidate.  It is possible for SuperB to search for light dark matter candidates by studying the decays of light mesons
$M$ to invisible final states~\cite{bevan:mcelrath}.  The light mesons discussed in the literature include $\Upsilon$,
$\eta$, $J/\Psi$. The SM process for these decays is $M\to \nu\nubar$. Additional amplitudes from NP, such as $M\to
\chi\chi$ could significantly enhance the observed branching fraction for such decays. For example the decay
$\Upsilon(1S)\to invisible$ has a SM branching fraction of $10^{-4}$, whereas new physics could enhance this up to the
current experimental limit from Belle of $2.5\times 10^{-3}$~\cite{bevan:belle1s}.  This limit used only $7\ifb$ of
data collected at the $\Upsilon(1S)$, whereas SuperB has the potential to collect hundreds of \ifb.  It would be
possible to measure the $\Upsilon(1S)\to invisible$ decay at the level of the SM with SuperB in order to place precise
constraints scenarios with light dark matter candidates.  Similar constraints will be possible using other $M\to
invisible$ decays.

\subsubsection{\boldmath{$\Delta S$ Measurements}}

It is possible to probe the presence of new physics in loops by comparing the tree level measurement of $\sin(2\beta)$
from $c\overline{c}s$ decays with the $\sin(2\beta_{\mathrm{eff}})$ measured from loop dominated $b\to s$ penguin, and
the loop and tree $b\to d$ transitions.  In order to correctly compare these quantities one has to compute the quantity
$\Delta S = \sin(2\beta_{\mathrm{eff}}) - \sin(2\beta) - \Delta S_{\mathrm{SM}}$.  Here quantity $\Delta
S_{\mathrm{SM}}$ accounts for higher order contributions to the difference coming from neglected standard model (SM)
processes. Some of the possible $\Delta S$ measurements are golden modes such as $\eta^\prime K^0$, $\phi K^0$, and
$K^+K^-K^0$ which all have small $\Delta S_{\mathrm{SM}}$. Theoretical uncertainties calculated for different $b\to s$
penguin decays by several groups are shown in Figure~\ref{fig:deltas}. The figure is divided into decay modes, and
each decay mode has up to four error bands drawn on it.  These error bands come from (top to bottom) calculations by
Beneke \etal~\cite{bevan:theory:beneke}, Williamson and Zupan~\cite{bevan:theory:williamson_zupan}, Cheng
\etal~\cite{bevan:theory:cheng}, and Gronau \etal~\cite{bevan:theory:gronau}. Figure~\ref{fig:deltas} summarizes
projections of current measurements of $\Delta S$ from the B-factories to 75\iab. Based on these projections the golden
modes have a $5\sigma$ discovery potential at SuperB.  Several other channels also have the potential for a $5\sigma$
discovery assuming that the theoretical uncertainties can be controlled at the level of a few percent.  When performing
these $\Delta S$ measurements we should also compare the values obtained with the results of theoretical predictions
based on clean interpretations of SM processes. It has recently been noted by Lunghi and Soni~\cite{bevan:soni} that by
comparing the loop to tree processes one is insensitive to possible new physics common to both, so it is important to
also compare the value of $\sin(2\beta_{(\mathrm{eff})})$ obtained in these measurements with the predictions of SM
based constraints for this observable. The data currently deviate by $2.1\sigma$ for the tree and $2.7\sigma$ for the
golden modes. SuperB is needed in order to determine if this is a first tantalizing hint of new physics of if this is
just a statistical fluctuation.


\begin{figure}[!ht]
\begin{center}
\resizebox{16cm}{!}{
\includegraphics{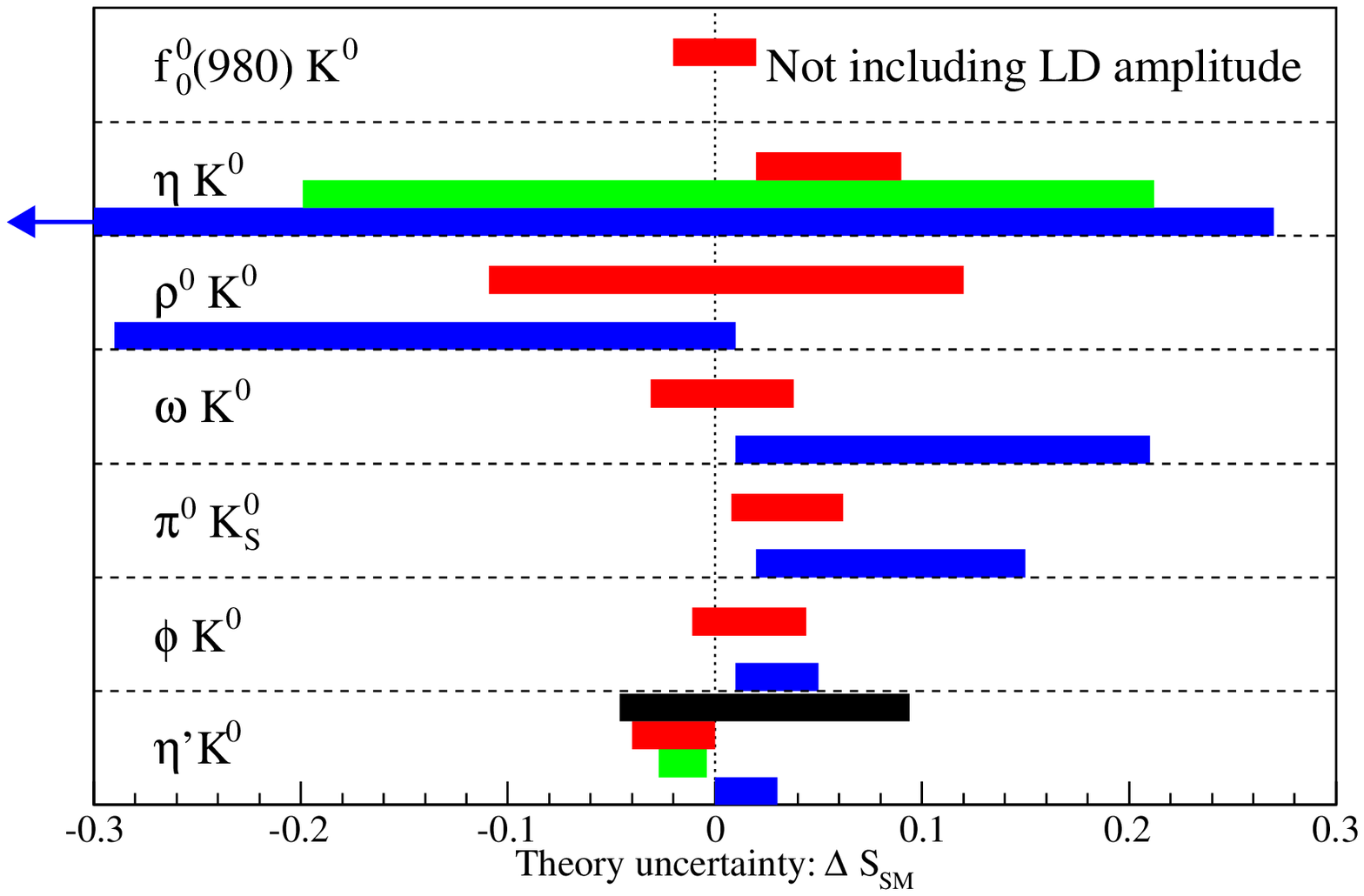}
\includegraphics{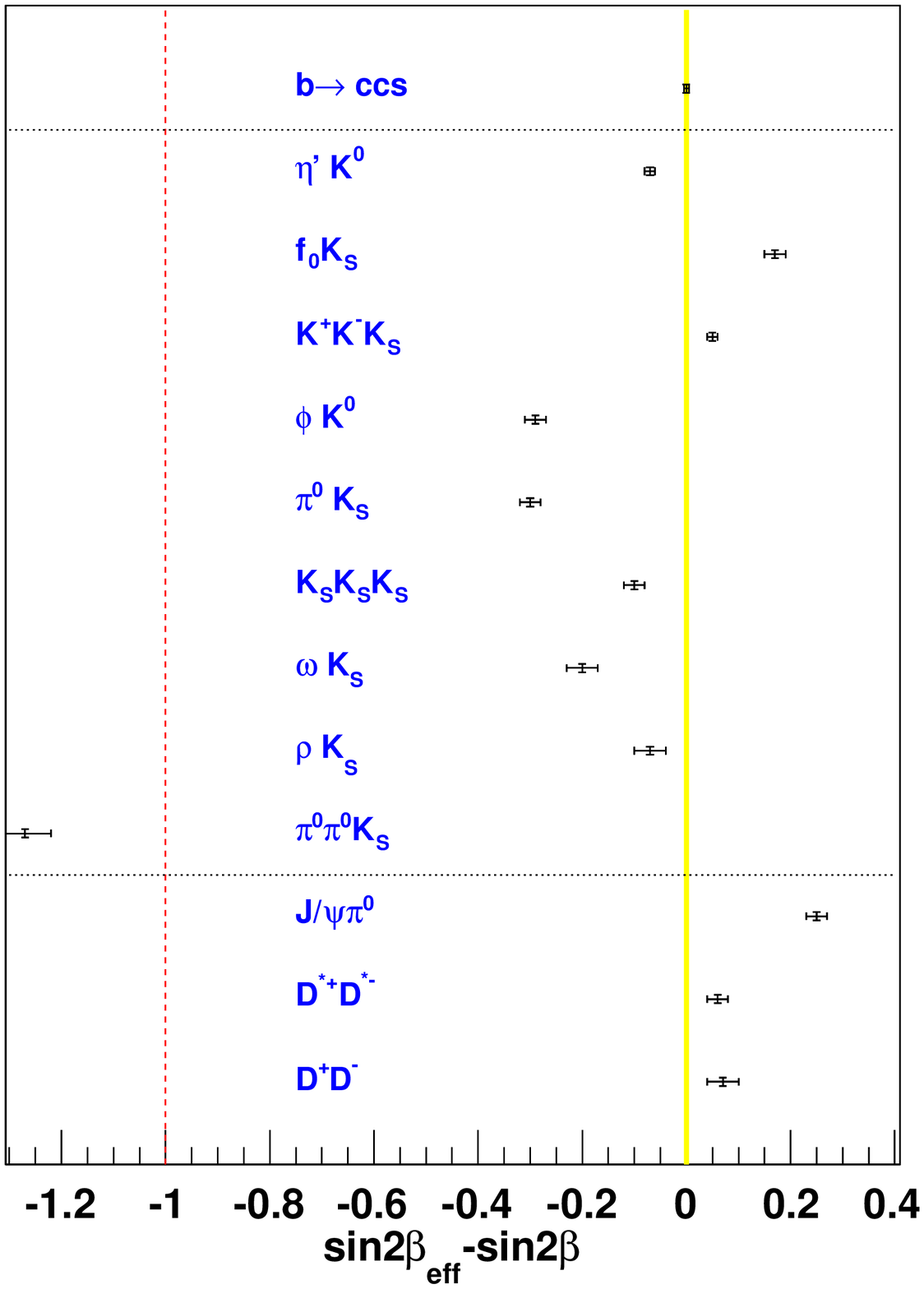}
}
\end{center}
\caption{(left) Theoretical uncertainties calculated for $\Delta S$ measurements of several $b\to s$ penguin processes
and (right) predictions for $\Delta S$ measurements with 75\iab of data at SuperB, extrapolating from current data.}
\label{fig:deltas}
\end{figure}

\subsection{Standard model measurements}

In addition to being able to search for and constrain new physics effects, SuperB will be able to perform many
precision tests of the SM.  In particular SuperB will be able to overconstrain the CKM matrix at the percent level
through measurements of the angles of the unitarity triangle.  Precise measurements of the sides of the triangle will
also be possible.  These tests of the SM will be able to limit the effects of new physics corrections to the quark
sector description of \CP violation in nature. One should keep in mind that while these precision over-constraints are
not the main aim of SuperB, they will be useful calibration measurements for the experiment.  In fact some of the
golden matrix measurements require this improved determination of SM parameters in order to elucidate NP. They will
also serve to reduce SM uncertainties on many new physics sensitive rare processes such as neutral and charged decays
$K\to \pi \nu\overline{\nu}$~\cite{bevan:augusto} which will be measured in the coming decade by the NA62 and KOTO
experiments. Details of the anticipated precision that SuperB can reach on SM calibration modes are given in the CDR
and Valencia workshop proceedings.

\subsection{Current Status of the Project}

The process of preparing a TDR has just begun.  It is anticipated that this will be completed by the end of 2010 or
early in 2011.  The TDR will contain a detailed description of the accelerator facility, including the accelerator
lattice description, as well as the detector and the main elements of the physics case for the SuperB experimental
programme.  Aspects of the detector design as described in the CDR will be revisited while fully optimizing SuperB for
a broad range of new physics searches.  This process will be done while keeping in mind that we intend to re-use the
DIRC quartz bars, CsI(Tl) calorimeter barrel, and super-conducting solenoid magnet from \babar. There is a lot of work
to do over the coming two years, however all of the necessary simulation, computing infrastructure and analysis tools
are available to complete this task.  It is anticipated that the SuperB complex will be constructed within five years
of TDR completion and approval of the project funding.  This timescale corresponds to data taking as early as mid-2015.

\section{Summary}

SuperB will be able to elucidate our understanding of new physics and the flavour problem by performing high precision
measurements irrespective of the results of the CERN particle physics programme.  In the event that new physics is
found at the LHC, SuperB will be able to constrain the flavour parameters that will be inaccessible to the CERN
experiments.  If no new physics is found at the LHC, the indirect energy scale search potential of SuperB will be
several hundred TeV, which is several orders of magnitude greater than any scale that can be directly accessed in any
existing or planned facility. Most of the golden matrix measurements at SuperB simply can not be made by the CERN
experimental programme, these include the $\tau$ lepton flavour violation and \CP violation measurements, most of the
$\Delta S$ measurements and the rare decays discussed in these proceedings.  Although the potential for charm physics
is not discussed in detail in these proceedings, there are many important measurements of charm decays that will be
done at SuperB.  These include searches for \CP violation and other fundamental tests of the SM that could be sensitive
to the effects of new physics.  The avid reader will find more details on the SuperB physics programme in the
Conceptual Design Report and Valencia Physics Workshop proceedings \cite{bevan:cdr,bevan:valencia}.

In order for this experiment to be a success, it is important to understand the golden matrix of physics observables
versus models that can be used to distinguish between different forms of new physics proposed in the literature.  By
mapping out the golden matrix, we can be sure that any future measurements found to deviate significantly from the SM
of particle physics can be used to test our theoretical understanding of high energy physics, and distinguish between
those hypotheses that remain compatible with measurement. Many signatures of physics beyond the SM have distinct
predictions that are only manifest through the parameters of broken flavour symmetry, and these parameters are best
accessed via an experiment like SuperB.  Measurements made in order to elucidate the flavour problem could have a
significant impact on the related universal asymmetry problem.  A TDR will be prepared over the next two years that
will refine the design of the SuperB facility and the ability to determine the observables required to distinguish
between NP scenarios in the golden matrix.  Once the TDR has been completed, it will take five years to construct the
SuperB facility (drawing from the experience of constructing \babar).  Based on these timescales SuperB could start to
take data as early as mid-2015.

\section{Acknowledgements}

The author would like to thank the meeting organisers for the opportunity to present this work at such an inspiring
conference, the SuperB community, and in particular Andy Wolski for useful discussions related to accelerator physics.

\end{document}
